\documentclass[journal=jctcce,manuscript=article,layout=twocolumn]{achemso}
\setkeys{acs}{usetitle=true}
\usepackage{graphicx}
\usepackage{subfigure}
\usepackage{amsmath}
\usepackage[version=3]{mhchem} 
\usepackage[T1]{fontenc}       

\title{Size-dependence of non-empirically tuned DFT starting points for $G_0W_0$ applied to $\pi$-conjugated molecular chains} 

\author{Juliana Bois}
\affiliation{Institut f\"ur Chemie, Universit\"at Potsdam, Karl-Liebknecht-Stra\ss{}e 24-25, 14476 Potsdam, Germany}

\author{Thomas K\"orzd\"orfer}
\email{koerz@uni-potsdam.de}
\affiliation{Institut f\"ur Chemie, Universit\"at Potsdam, Karl-Liebknecht-Stra\ss{}e 24-25, 14476 Potsdam, Germany}


\date{\today}
\begin{document}
\begin{tocentry} 
\begin{center}
\includegraphics[width=1.0\textwidth]{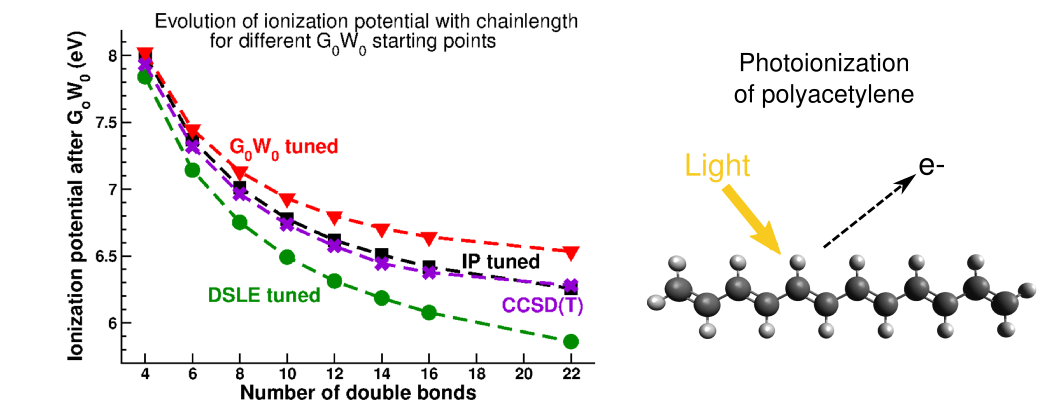}
\end{center}
\end{tocentry}

\begin{abstract}
$G_0W_0$ calculations for predicting vertical ionization potentials (IPs) and electron affinities of molecules and clusters are known to show a significant dependence on the density functional theory (DFT) starting point. A number of non-empirical procedures to find an optimal starting point have been proposed, typically based on tuning the amount of HF exchange in the underlying hybrid functional specifically for the system at hand. 
For the case of $\pi$-conjugated molecular chains, these approaches lead to a significantly different amount of HF exchange for different oligomer sizes. 
In this study, we analyze if and how strongly this size dependence affects the ability of non-empirical tuning approaches to predict accurate IPs for $\pi$-conjugated molecular chains of increasing chain length. To this end, we employ three different non-empirical tuning procedures for the $G_0W_0$ starting point to calculate the IP of polyene oligomers up to 22 repeat units and compare the results to highly accurate coupled-cluster calculations. We find that, despite its size dependence, using an IP-tuned hybrid functional as a starting point for $G_0W_0$ yields excellent agreement with the reference data for all chain lengths. 
\end{abstract}

\maketitle

\section{Introduction}
An accurate prediction of the ionization potential (IP) of $\pi$-conjugated polymers is of utmost importance when considering them for applications in organic electronics. 
Due to their huge system size, accurate theoretical methods can not be used to investigate $\pi$-conjugated polymers directly. 
To circumvent this problem, relevant properties are often calculated for oligomers of increasing length and extrapolated to the infinite chain limit, therefore allowing the use of electronic structure methods such as density functional theory (DFT).
This is known as the oligomer approach, which is widely followed for the prediction of polymer properties.\cite{adv_mat_2007_opt_bandg,jpc_a_2012_rev_fits,acta_polym_1997_effective_conj_length}\\
Many body perturbation theory within the $GW$ approximation\cite{GW,GW_2,GW_3} is a method increasingly applied to calculate charged excitation energies, that is, IPs and electron affinities (EAs) of molecules.\cite{EPL2.298.2006,phys_rev_b_2011_gw_ip_organic,phys_rev_b_2012_gw_hierachy_results,GW100,GW100_planewave,GW100_scGW,PhysRevB.86.041110,PhysRevLett.86.472,PhysRevB.83.115123,j_phys_2017_GW_organic_semi,doi:10.1021/acs.jctc.5b01238} 
It can be applied using either full\cite{EPL2.298.2006,phys_rev_b_2010_gw_sc_on_dft,PhysRevB.90.085141,PhysRevB.86.081102,doi:10.1021/acs.jctc.5b01238} or partial\cite{GW_2,phys_rev_b_2011_gw_ip_organic,PhysRevB.83.115123, PhysRevLett.96.226402,PhysRevB.54.8411,doi:10.1021/acs.jctc.5b01238} self consistency, or as a single step 
correction on top of a DFT or Hartree-Fock (HF) calculation, hereafter referred to as $G_0W_0$. While fully self-consistent $GW$ calculations ($scGW$) are still scarce,
non-self-consistent $G_0W_0$ is by far the most commonly used variant, as it has been shown to lead to accurate IPs at a fraction of the cost of partially or fully self-consistent schemes.\cite{phys_rev_b_2010_gw_sc_on_dft,phys_rev_b_2012_gw_hierachy_results,phys_rev_b_2015_g0w0_tuning_polyacetylene}
However, while the $G_0W_0$ correction can be readily applied to any DFT calculation, a strong starting point dependence has been found,\cite{NJPhys_GW_OEP_Rinke_2005, PhysRevB.83.115123, PhysRevB.86.081102, PhysRevB.76.115109, phys_rev_b_2012_gw_hierachy_results, jctc_2013_gw_starting_points_benchmark, phys_rev_b_2011_gw_ip_organic, PhysRevB.86.041110,GW100,GW100_planewave,GW100_scGW,gallandi_koerzd_lrc_g0w0_2015,GW_acenes, jctc_2016_IP_III_GW_stuff,j_phys_2017_GW_organic_semi,phys_rev_b_2011_gw_copper,doi:10.1021/acs.jctc.5b01238} hence making the choice of the exchange-correlation functional in the underlying DFT calculation crucial for the accuracy of the calculated IPs and EAs.

The starting point dependence of $G_0W_0$ has been connected to mainly two effects. First, $G_0W_0$ is based on a perturbative correction, so it can only be expected to yield reliable results if the eigenvalues of the DFT starting point are already good approximations to their corresponding quasiparticle energies. Second, the validity of the quasiparticle approach hinges on a correct description of the screened interaction $W$, which is mainly determined by the systems polarizability and, hence, its band gap. Consequently, it is essential that the eigenvalue gap between the highest occupied molecular orbital (HOMO) and the lowest unoccupied molecular orbital (LUMO) of the DFT starting point approximates the true quasiparticle gap, i.e., the difference between the IP and EA.\cite{j_phys_2017_GW_organic_semi} 

Semilocal exchange-correlation functionals fulfill neither of these two requirements, which is why using hybrid functional starting points that include a fraction of HF exchange has been found to be essential for obtaining accurate IPs and EAs from $G_0W_0$ calculations.\cite{PhysRevB.76.115109, PhysRevB.86.041110, jctc_2016_IP_III_GW_stuff, phys_rev_b_2012_gw_hierachy_results,  GW100, GW100_planewave, GW100_scGW, gallandi_koerzd_lrc_g0w0_2015, GW_acenes} In recent years, a number of different non-empirical tuning schemes have been proposed, aiming to find an optimal amount of HF exchange for the $G_0W_0$ starting point.

One of the most promising non-empirical tuning approaches is the so-called IP-tuning,
\cite{IPtuning_JACS2009,jcp_2009_ct_excit_tuning, ann_rev_phys_chem_2010_tuning_RSH, pccp_2009_tuning_proc_paper}
which will be explained in more detail below. Due to their high accuracy for the prediction of IPs, EAs, and fundamental gaps, IP-tuned long-range corrected (LRC) hybrid functionals had long been suggested as a potentially interesting starting point for $G_0W_0$.\cite{PhysRevLett.109.226405} Recently, it was demonstrated that, indeed, the combination of these functionals with a single step $G_0W_0$ correction produces excellent results for the IPs and EAs of a organic molecules, clearly outperforming typically used semilocal or global hybrid starting points.\cite{gallandi_koerzd_lrc_g0w0_2015,jctc_2016_IP_II_lrc_funct_GW,jctc_2016_IP_III_GW_stuff, GW_acenes} In addition, IP-tuned hybrid functionals have been shown to be appropriate starting points for excited state $GW$ Bethe-Salpeter equation calculations.\cite{JChemPhys.146.194108}

However, the combination of the IP-tuning procedure with the oligomer approach has been shown to lead to several problematic shortcomings,\cite{jcp_2012_wPBE_PBEh_BLA_MESIE,jctc_2016_BLA_Ex_thermal,acc_chem_res_2014_tuning_polyene} which can be traced back to the fact that IP-tuned functionals are size dependent.\cite{jcp_2011_range_sep_conj, JChemPhys_Kuemmel_13, phys_rev_let_2016_tuning_lrc_polyene_baer_paper, jcp_2015_piecewise_linearity_solid_state_finite}
Especially for highly conjugated molecular chains, such as the ones frequently used in organic electronics and photovoltaics, the amount of exact exchange obtained by the IP-tuning procedure
(accounted for by the range-separation parameter $\omega$ in LRC functionals or the amount of HF exchange $\alpha$ in global hybrid functionals) decreases rapidly as the oligomer chain length increases, before gradually leveling off.\cite{jcp_2011_range_sep_conj,JChemPhys_Kuemmel_13, acc_chem_res_2014_tuning_polyene, jcp_2012_wPBE_PBEh_BLA_MESIE, phys_rev_let_2016_tuning_lrc_polyene_baer_paper, AccChemRes_Autschbach14, jcp_2015_piecewise_linearity_solid_state_finite} 
For
highly conjugated systems, convergence can only be found for oligomer chains of several nanometers.\cite{phys_rev_let_2016_tuning_lrc_polyene_baer_paper} 
This strong size-dependence has been demonstrated to create unphysical trends for properties that depend heavily on the amount of exact exchange, such as bond-length alternation\cite{jcp_2012_wPBE_PBEh_BLA_MESIE} and optical band-gaps.\cite{jctc_2016_BLA_Ex_thermal} 
The aim of this work is to investigate how this size dependence affects the reliability of non-empirically tuned hybrid functionals as a starting point for $G_0W_0$ when calculating the IPs of $\pi$-conjugated polymers of increasing chain length.

\begin{figure}[tbh]
\includegraphics[width=2.5cm]{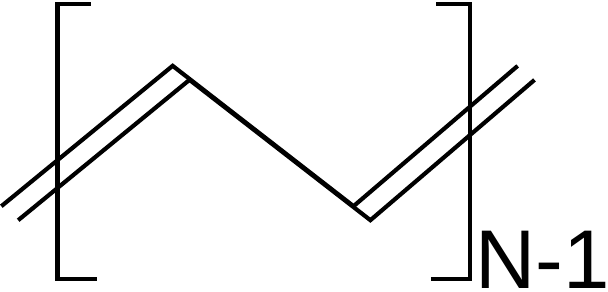}
\caption{Structure of trans-polyacetylene, where N equals the number of double bonds.\label{polyacetylene}}%
\end{figure}
Polyacetylene (H$-$(HC$=$CH)$_N-$H, see Fig.~\ref{polyacetylene}) serves as an ideal model for $\pi$-conjugated polymers. Its simple structure and small repeat units allow calculations for oligomers with many repeat units using highly accurate wavefunction methods such as coupled cluster singles, doubles, perturbative triples (CCSD(T)), thus, providing a reliable theoretical reference for computationally more efficient methods such as DFT and/or $G_0W_0$. At the same time, it shows the $\pi$-conjugated backbone found in all polymers of this class, exhibiting the typical pattern of single and double bonds. \cite{phys_rev_let_1979_solitons_polyene,jcp_1985_polyene_oligomers_geom_elec,jpc_a_2001_ex_tddft_polyene_oligomers} 
The difference in bond length between single and double bonds, known as bond length alternation (BLA), is a defining property of $\pi$-conjugated polymers. Predicting the BLA and its evolution with chain-length is a notorious problem many standard computational methods struggle with. \cite{jcp_1997_BLA_polyacetylene_dft,jpc_a_2006_BLA_hybrid_functionals,jctc_2011_dft_wavefun_benchmarks_BLA,jcp_2012_wPBE_PBEh_BLA_MESIE}
This problem is particularly severe since the BLA decisively influences the electronic and optical properties of organic $\pi$-conjugated molecular chains.

In the following methodology section we will further elaborate on the aforementioned difficulties of common density functionals to act as a reliable starting point for $G_0W_0$ and provide the theoretical background for the three non-empirical tuning procedures studied in this work.
In the results section, we will then compare the $\Delta$SCF and HOMO IPs 
obtained from each tuning method to CCSD(T) reference calculations for the oligomers of polyacetylene up to a chain length of N=22. 
Finally a critical evaluation of the tuning methods' applicability as a reliable starting point for $G_0W_0$ as well as their respective behavior with increasing oligomer size will follow.

\section{Methodology}
The aim of this section is to introduce the methods used in this article, especially highlighting the differences in the non-empirical tuning procedures.
\subsection*{The $\boldsymbol{G_0W_0}$ method}
$GW$ is an approximation made in many-body perturbation theory, expressing the self-energy as a first order expansion in terms of the single particle Greens function, $G$, and the screened Coulomb interaction, $W$.\cite{GW,GW_2,GW_3} 
Fully self-consistent $GW$ is computationally unviable for medium/large systems and results in comparatively large errors for IPs of organic molecules.\cite{jctc_2016_IP_III_GW_stuff, phys_rev_b_2012_gw_hierachy_results, GW100_scGW} 
In contrast, carrying out a single step perturbative correction on top of a DFT calculation, as done in $G_0W_0$ calculations, is significantly cheaper and has been shown to give excellent results for IPs when combined with a well-considered 
starting point.\cite{phys_rev_b_2010_gw_sc_on_dft,phys_rev_b_2012_gw_hierachy_results,jctc_2016_IP_II_lrc_funct_GW, jctc_2016_IP_III_GW_stuff, GW100, GW100_scGW, gallandi_koerzd_lrc_g0w0_2015, GW_acenes, PhysRevB.86.041110, phys_rev_b_2016_dsle_tuning, phys_rev_b_2013_g0w0_tuning,phys_rev_b_2015_g0w0_tuning_polyacetylene}  
The $G_0W_0$ quasiparticle energies are given by
\begin{equation} \label{eq:g0w0}
E_i^{\mathrm{QP}}={\epsilon}_{i}^{\mathrm{DFT}} + \langle\phi_i^{\mathrm{DFT}}| \hat{\sum}(E_i^{\mathrm{QP}})-\hat{v}_{xc}|\phi_i^{\mathrm{DFT}}\rangle
\end{equation}
where ${\epsilon}_{i}^{\mathrm{DFT}}$ are the underlaying DFT eigenvalues, $\phi_i^{\mathrm{DFT}}$ are the corresponding orbitals, $\hat{\sum}(E_i^{\mathrm{QP}})$ is the $GW$ self-energy calculated using the DFT
orbitals and eigenvalues, and $\hat{v}_{xc}$ is the exchange-correlation operator.\cite{notevxc}
As the eigenvalues and orbitals of the DFT calculation to which the $G_0W_0$ correction is applied are directly used
as input for the correction, it is vital that these are good approximations to actual quasiparticle energies.

\subsection*{Deviation from straight line}
\begin{figure}[t]
\includegraphics[width=1\linewidth]{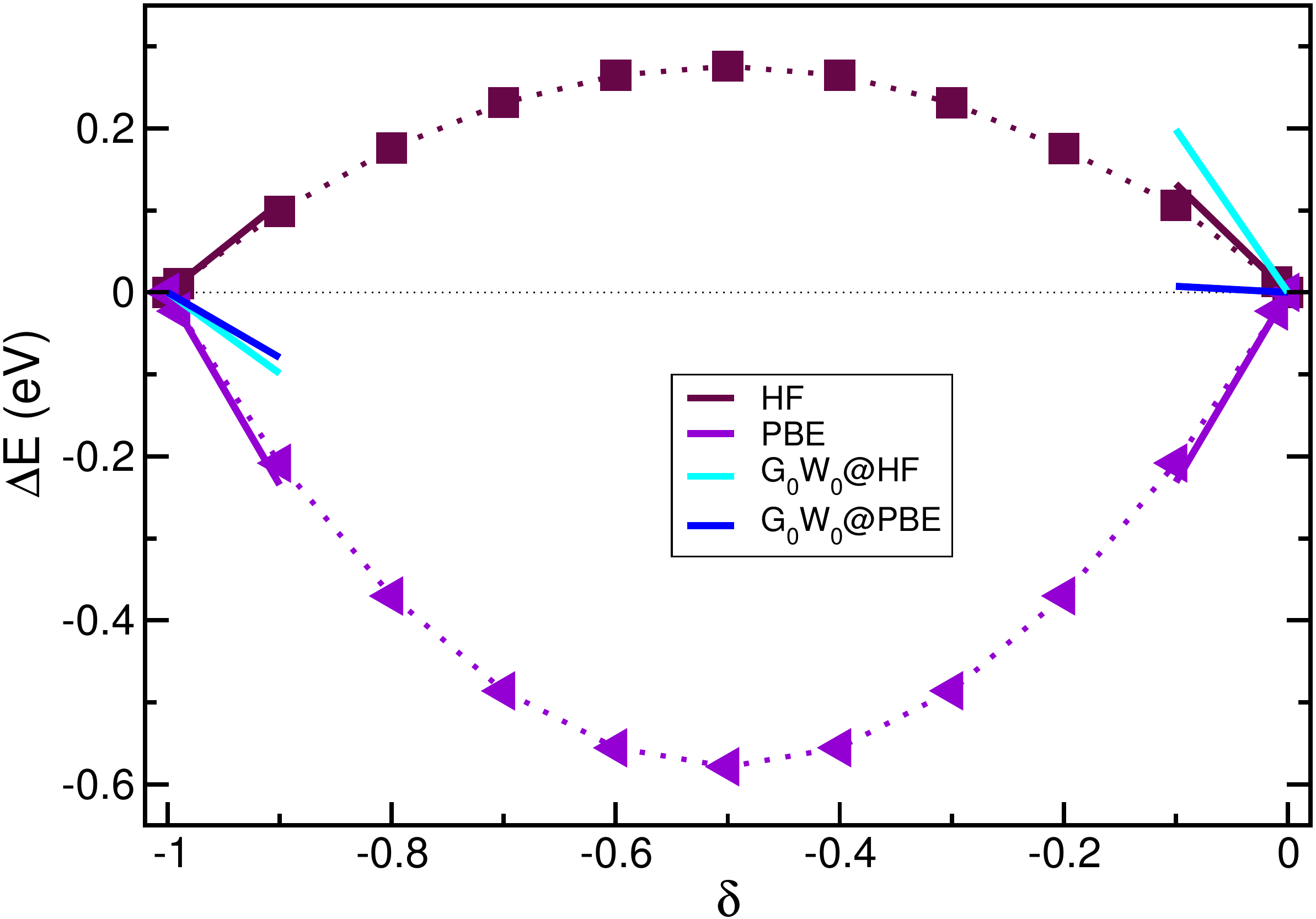}
\caption{Energy deviation from straight line behavior calculated using PBE and HF for Octatetraene (N=4). Lines show the initial and final slopes for PBE and HF before and after a $G_0W_0$ correction, 
given by the cation LUMO and neutral HOMO eigenvalues.\label{DSLE_frac}}
\end{figure}

The exact total energy of any many-electron system exhibits straight line behavior when going from M electrons to M$\pm$1 electrons via fractional electron numbers ($\delta$), while it shows a derivative discontinuity at integer electron numbers.\cite{phys_rev_let_1982_derivative_discont} The difference in slope when going from M-1 to M and from M to M+1 electrons equals the difference between the systems' IP and EA. 
Standard DFT functionals, however, have been shown to not reproduce this behavior. \cite{jcp_2006_mesie_common_functionals,jcp_2006_mesie_frac_charge_dft_functionals, phys_rev_let_2008_loc_deloc_error_bandgaps} Fig.~\ref{DSLE_frac} shows the deviation from straight line behavior for fractional electron numbers for Octatetraene (N=4) obtained from
a semilocal functional and the HF approach. Semilocal functionals, represented here by the functional of Perdew-Burke-Ernzerhof (PBE), show no derivative discontinuity and exhibit a convex slope between occupation numbers, thereby artificially lowering the
energy for fractional electron numbers. This is known as the {\sl delocalization error} of semilocal DFT. In contrast, the HF approach shows an exaggerated derivative discontinuity and a concave slope between the M and M$\pm$1 electron systems. This strongly
stabilizes the energy of integer electron occupations, causing what is known as the {\sl localization error}.
The integral of the curves in Fig.~\ref{DSLE_frac} is often referred to as the many-electron self interaction error (MESIE).\cite{jcp_2006_mesie_frac_charge_dft_functionals,jcp_2006_mesie_common_functionals}

As calculations with fractional electron numbers are both time consuming and, more importantly, not possible for $GW$, 
it can be advantageous to follow an alternative method to quantify the curvature of the energy.\cite{PhysRevLett.103.176403,PhysRevB.81.205118,phys_rev_b_2016_dsle_tuning} 
This deviation from straight line error (DSLE) is defined as the energy difference between the EA of the cationic system, $\mathrm{EA_c}(M-1)$, and the IP of the neutral system, $\mathrm{IP}(M)$. 
\begin{equation} \label{eq:dsle_general}
\begin{split}
{\mathrm{DSLE}} &= \mathrm{EA}(M-1) - \mathrm{IP}(M) \\
    &= \epsilon_{\mathrm{LUMO,cat}} - \epsilon_{\mathrm{HOMO,neut}} \, .
\end{split}
\end{equation}
Within the framework of DFT,  $\mathrm{EA_c}(M-1)$ and $\mathrm{IP}(M)$ are represented by the LUMO eigenvalue of the cation, $\epsilon_{\mathrm{LUMO,cat}}$, and HOMO eigenvalue of the neutral molecule, $\epsilon_{\mathrm{HOMO,neut}}$, respectively. 
This definition of the DSLE relies on the fact that, according to Janak's theorem,\cite{Ja78} the energy slope of the fractionally occupied HOMO is given by the HOMO eigenvalue, i.e.,
\begin{equation} \label{eq:janak}
\frac{{\delta}E}{{\delta}f_{\mathrm{HOMO}}}={\epsilon}_{\mathrm{HOMO}}\, .
\end{equation}
Consequently,
the initial and final slopes of the total energy curve are given by the cations' LUMO ($\delta = -1$) and the neutral molecules' HOMO ($\delta = 0$) eigenvalue, respectively. In Fig.~\ref{DSLE_frac}, these eigenvalues are represented by short lines with the corresponding slopes, showing excellent agreement with the calculated full total energy difference curves.
A table showing the general agreement between the DSLE found for initial/final slopes of fractional electron numbers and the cation LUMO and the neutral HOMO eigenvalues can be found in the supporting information. 
We therefore conclude that calculating the DSLE based on the neutral molecules' HOMO and the cations' LUMO eigenvalues provides for a reliable measure of the deviation from linearity without having to go through the numerically expensive fractional electron number calculations.

\subsection*{IP tuning}
IP-tuning is a well established non-empirical tuning procedure based on enforcing 
the functional to obey the IP-theorem,\cite{IPtuning_JACS2009,jcp_2009_ct_excit_tuning, ann_rev_phys_chem_2010_tuning_RSH, pccp_2009_tuning_proc_paper} i.e.,
\begin{equation} \label{eq:ip_theorem}
\mathrm{IP}=E_M - E_{M-1}=-\epsilon_{\mathrm{HOMO}}.
\end{equation}
This fixes the slope of the fractional electron number curve in Fig.~\ref{DSLE_frac} to be zero at $\delta=0$. Consequently, IP-tuning ensures not only the IP-Theorem but also (approximately) the straight-line behavior of the total energy curve when going from the neutral molecule to the cation.
Global hybrid functionals have been previously discussed in the literature in the context of IP-theorem based tuning procedures.\cite{IMAMURA2011130,PhysRevLett.106.226403} 
Here, we have used the global hybrid functional PBEh for all tuning, which combines PBE with a fraction of exact exchange:
\begin{equation} \label{eq:pbeh}
 E_{xc}^{\mathrm{PBEh,\alpha}}= \alpha E_x^{\mathrm{HF}} + (1-\alpha) E_x^{\mathrm{PBE}} + E_c^{\mathrm{PBE}}.
\end{equation}
The IP-tuned fraction of exact exchange, $\alpha$, is found by minimizing 
\begin{equation} \label{eq:ip_tuning}
\Delta_{\mathrm{IP}} = |-\epsilon^{\alpha}_{\mathrm{HOMO,neut}}-(E^{\alpha}_{\mathrm{neut}}-E^{\alpha}_{\mathrm{cat}})|
\end{equation}
with $E^{\alpha}_{\mathrm{neut/cat}}$ representing the energy of the neutral and cationic systems calculated on the neutral geometry and $\epsilon^{\alpha}_{\mathrm{HOMO,neut}}$ is
the HOMO eigenvalue of the neutral molecule.

IP-tuning is perhaps the most well known non-empirical tuning procedure since it has been extensively discussed in the literature and used for a wide range of applications.\cite{jpc_a_2011_ip_tuning_metal_complex,jctc_2011_tuning_pi_olig, phys_rev_let_2011_fund_gap_tuning, jctc_2012_optical_rot_tuning, phys_rev_b_2012_tuning_qp_spec, acc_chem_res_2014_tuning_polyene,AccChemRes_Autschbach14, ann_rev_phys_chem_2010_tuning_RSH, phys_rev_b_2011_range_sep_size_dependent, JCP134.151101.2011, jcp_2011_triplet_instabilities, jcp_2014_torsion_potentials, JCTC5.1934.2014, JPCC8.3925.2014,PhysRevB.88.081204,PhysRevLett.106.226403}
Importantly, IP-tuning has originally been introduced and is, indeed, most frequently used as a computationally efficient alternative to $GW$ calculations, since it has been demonstrated to provide for very accurate IPs and EAs. In a recent theoretical benchmark study on a test set of organic acceptor molecules\cite{jctc_2016_accurate_IP_I_ccsdt_reference, jctc_2016_IP_II_lrc_funct_GW, jctc_2016_IP_III_GW_stuff} as well as in a study that directly compared to highly resolved gas-phase photoelectron spectroscopy measurements,\cite{gallandi_koerzd_lrc_g0w0_2015} IP-tuned LRC hybrid functionals outperformed $scGW$ and $G_0W_0$ calculations based on commonly used semilocal DFT or HF starting points. 
However, in the same studies it was demonstrated that the results of the IP-tuned functionals could be further improved by applying a $G_0W_0$ correction to the DFT eigenvalues. 
Here, we will study the performance of IP-tuned global hybrid functionals both with and without a $G_0W_0$ correction, where a special emphasis lies on the size dependence of these functionals and the potential consequences this may have for the application to long, $\pi$-conjugated molecular chains.

\begin{figure}[tb]
\includegraphics[width=1\linewidth]{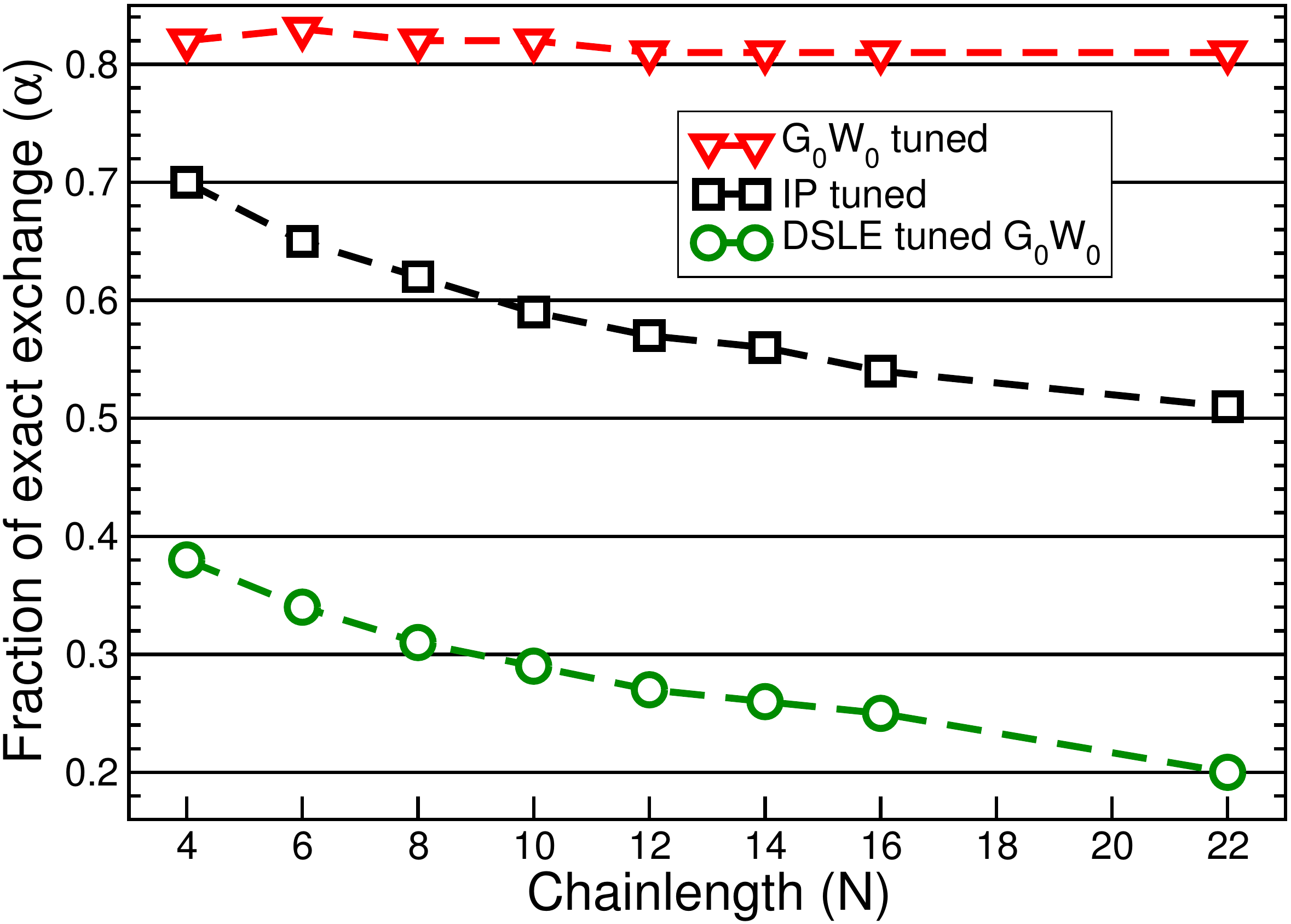}
\caption{Amount of exact exchange found through IP-tuning, $G_0W_0$-tuning and DSLE tuning for polyacetylene oligomers of increasing size with PBEh* geometries.\label{tuning}}
\end{figure}

The system size dependence of the IP-tuned exact exchange parameter can be clearly observed when IP-tuning PBEh for the oligomers of polyacetylene. As demonstrated in Fig.~\ref{tuning}, the IP-tuned fraction 
of exact exchange drops off as the chain length increases; starting at 0.70 for N=4 it drops to 0.51 for N=22. While a slight sloping off can be observed towards longer chain lengths, no convergence can be observed for the oligomer sizes studied here. Since a different amount of HF exchange and, hence, a different exchange-correlation functional is used for different oligomer lengths, 
the IP-tuning procedure shows strong size dependence and is therefore likely to fail in predicting accurate values for the polymer limit via extrapolation of the results obtained from the individual oligomers. 

\subsection*{DSLE tuning}
Along with the fractional particle curves and DFT eigenvalue slopes, Fig.~\ref{DSLE_frac} also shows the
slopes derived from the corresponding quasiparticle energies $\epsilon^{\mathrm{QP}}_{\mathrm{LUMO,cat}}$ and $\epsilon^{\mathrm{QP}}_{\mathrm{HOMO,neut}}$ obtained from a $G_0W_0$ correction to the PBE and HF eigenvalues, respectively. This highlights that, depending on the starting point, a significant DSLE may still remain after the $G_0W_0$ correction. In other words, $G_0W_0$ based on a PBE or HF starting point is not free from DSLE. 

This finding has recently lead {\sl Rinke et al.} to propose an alternative non-empirical tuning procedure for the DFT starting point of $G_0W_0$.\cite{phys_rev_b_2016_dsle_tuning}
While the IP-tuning procedure significantly reduces the DSLE in the DFT starting point, the central idea of the DSLE tuning is to enforce the straight line behavior {\sl after} the $G_0W_0$ correction.
This is achieved by tuning the amount of HF exchange in the PBEh starting point in order to minimize
\begin{equation} \label{eq:dsle_tuning}
\begin{split}
\Delta_{\mathrm{DSLE}} &= \mathrm{EA}(M-1) - \mathrm{IP}(M) \\
    &= \epsilon^{\alpha,\mathrm{QP}}_{\mathrm{LUMO,cat}} - \epsilon^{\alpha,\mathrm{QP}}_{\mathrm{HOMO,neut}}\, ,
\end{split}
\end{equation}
where EA$_c(M-1)$ is the EA of the cationic system
and IP$(M)$ is the IP of the neutral system.
It is important to remember that, when following the DSLE-based tuning to minimize Eq.~(\ref{eq:dsle_tuning}), the eigenvalues considered are not the DFT eigenvalues but the quasiparticle energies obtained from 
the $G_0W_0$ correction.

When tuning $G_0W_0$ on PBEh for a series of polyacetylene oligomers of increasing length, a decrease in $\alpha$ with increasing chainlength similar to IP-tuning can be observed for the DSLE tuned functional (see Fig.~\ref{tuning}). 
The fractions of exact exchange are, however, much lower for all chain length, with an $\alpha$ of 0.38 for N=4 which falls off to 0.20 for N=22. 

\subsection*{$\boldsymbol{G_0W_0}$ tuning}
While both tuning procedures introduced above are constructed around the idea of enforcing straight line behavior, {\sl Scheffler et al.} came up with an entirely different approach to non-empirically tune the 
fraction of exact exchange in a hybrid functional.\cite{phys_rev_b_2013_g0w0_tuning,phys_rev_b_2015_g0w0_tuning_polyacetylene} Their tuning procedure is based on the idea of minimizing the single-shot $G_0W_0$ correction to the HOMO energy of a DFT calculation.
Following Eq.~(\ref{eq:ip_theorem}), the HOMO energy of a system is strictly equal to the IP for the exact functional, meaning the DFT calculation describes the systems' IP accurately and 
the $G_0W_0$ correction should vanish. This is, however, not true for common functionals, where the correction found from $G_0W_0$ can be significant.\cite{phys_rev_b_2010_gw_sc_on_dft,phys_rev_b_2011_gw_ip_organic,phys_rev_b_2015_g0w0_tuning_polyacetylene}
By adapting the fraction of exact exchange in the DFT starting point, $G_0W_0$ tuning aims to minimize the self-energy $G_0W_0$ correction to the HOMO IP, i.e., 
\begin{equation} \label{eq:gw_tuning}
\Delta_{G_0W_0} = |\epsilon^{\alpha,\mathrm{DFT}}_{\mathrm{HOMO}}-\epsilon^{\alpha,\mathrm{QP}}_{\mathrm{HOMO}}|
\end{equation}
where $\epsilon^{\alpha,\mathrm{DFT}}_{\mathrm{HOMO}}$ is the DFT eigenvalue and $\epsilon^{\alpha,\mathrm{QP}}_{\mathrm{HOMO}}$ is the $G_0W_0$ quasiparticle energy of the HOMO. When compared to the other tuning methods, one benefit of $G_0W_0$ tuning is that it only requires calculations on the neutral molecule, not the cation. Similar to the DSLE tuning, it requires several $G_0W_0$ calculations to find the tuned $\alpha$ value. In contrast, IP-tuning only requires computationally more efficient DFT calculations for determining $\alpha$.

Different from the other tuning methods, the fraction of exact exchange found for polyacetylene oligomers using $G_0W_0$ tuning shows no clear trend with oligomer length, see Fig.~\ref{tuning}. Additionally, the tuned $\alpha$ values are significantly lager than for the other tuning methods.
While a very slight fluctuation between $\alpha$=0.82 and 0.83 can be observed from N=4 to N=10, $\alpha$ stabilizes at 0.81 for N$\geq$12. This points towards the $G_0W_0$ tuned functionals being
approximately size-consistent, at least for the particular case of the polyene oligomers studied in this work. 

\subsection*{Oligomer approach and size dependence}
It has been shown that IP tuning faces difficulties when used to calculate properties of polyacetylene oligomers of increasing length. As the spatial extension of the $\pi$-conjugated system increases, the amount of exact exchange required to minimize the error in Eq.~(\ref{eq:ip_tuning}) falls off considerably.\cite{jcp_2011_range_sep_conj, phys_rev_let_2016_tuning_lrc_polyene_baer_paper} This has important consequences for the calculation of structural, electronic, or optical properties.
This is exemplified by the example of polyacetylene, where the amount of BLA strongly depends
on the amount of exact exchange present in a functional used for the structure optimization.\cite{jcp_2012_wPBE_PBEh_BLA_MESIE,jctc_2016_BLA_Ex_thermal}
While the extrapolation of CCSD(T) results predicts convergence of the BLA to $\sim$0.08\,{\AA} after about 8-10 repeat units, IP-tuned functionals show a continued dropping off for increasing oligomer lengths, resulting in grossly underestimated BLA for long chains.\cite{jcp_2012_wPBE_PBEh_BLA_MESIE,acc_chem_res_2014_tuning_polyene} A similarly erroneous behavior 
is encountered for optical excitation
energies, which were also shown to depend heavily on the amount of HF exchange, independently of any geometry effects.\cite{jctc_2016_BLA_Ex_thermal}

To understand the origin of this problem, it is helpful to consider the change of the systems energy upon addition or removal of an electron in the limit of an infinite chain. Semilocal functionals exhibit a strong delocalization error, which means 
it is energetically favorable for an electron (or hole) to delocalize over the entire $\pi$-conjugated system. 
For an infinitely long molecular chain this means the electron would delocalize over an infinite number of atoms, such that only an infinitesimally small fraction of an electron is added to each repeat unit. As a consequence, the energy no longer curves between $M$ and $M\pm1$ but instead behaves like the initial slope of the curve for all fractions of electron added/removed, resulting in the linearity of the total energy between the $M$ and $M\pm1$ electron systems.\cite{phys_rev_let_2008_piecewise_linearity_finite_infinite,jcp_2015_piecewise_linearity_solid_state_finite} It should
be noted, however, that although the MESIE vanishes, a delocalization error is still very much present but simply a lot harder to quantify.
Another consequence is that the IP-theorem no longer bears significant meaning, as every approximate exchange-correlation functional will give a straight line between $M$ and $M-1$ and, consequently, obey the IP-Theorem. Still, each functional might yield a considerably different IP. Consequently, the IP-tuning procedure can not straightforwardly be applied to infinite systems. 

\begin{figure}[tb]
\includegraphics[width=1\linewidth]{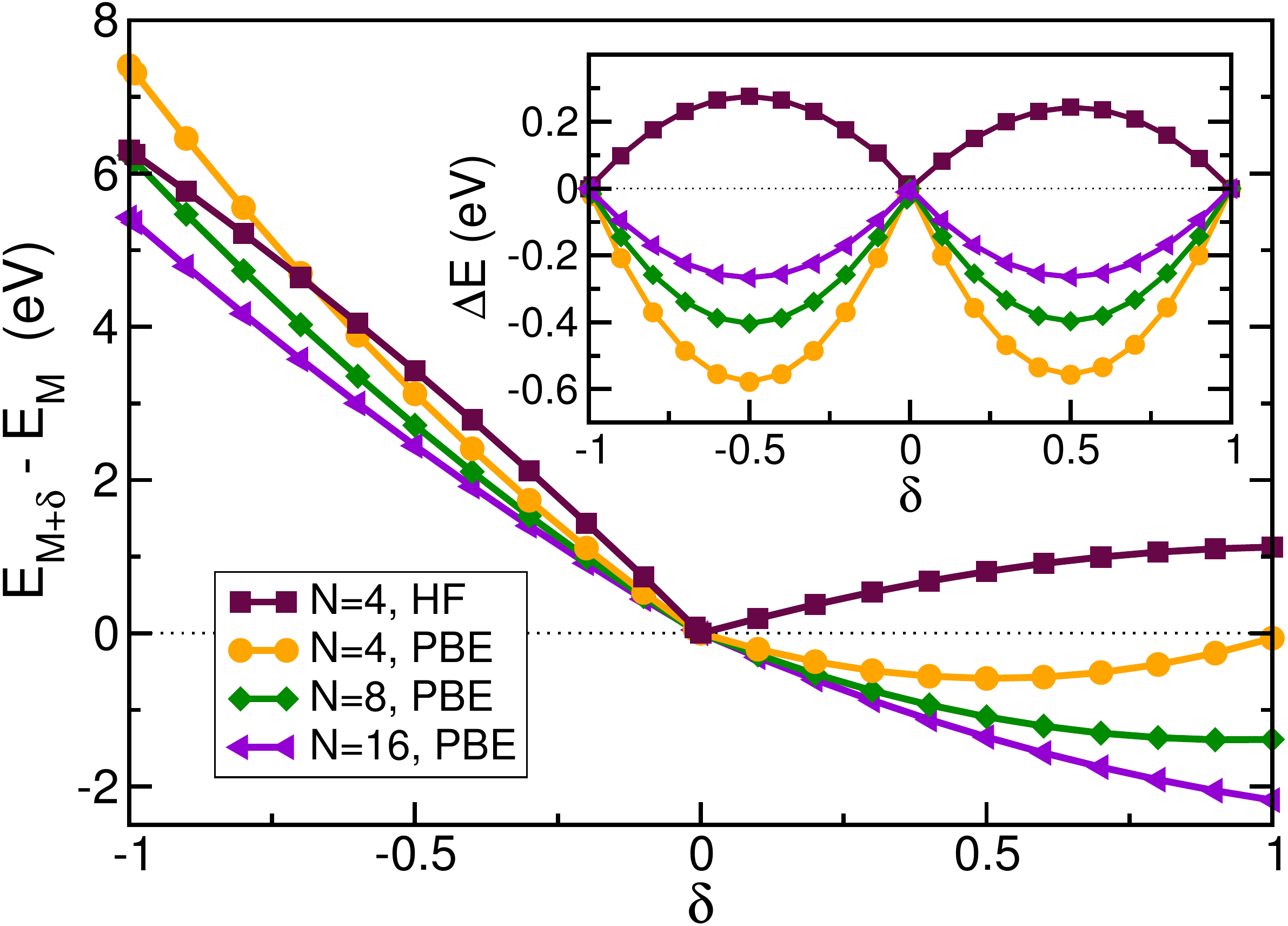}
\caption{Change of energy for fractional occupation numbers for polyacetylene oligomers of increasing length, where N is the number of double bonds. For a semilocal functional such as PBE, appearance of piecewise
linearity can be observed with increasing system size, leading to a decrease of the MESIE (see inset).\label{MESIE_frac}}
\end{figure}

While this argument only strictly holds for infinitely large molecular chains, it has been shown that it has important consequences also for finite molecular chains of increasing size. As the system size increases, the delocalization
error found for semilocal DFT steadily decreases as the energy curve for fractional occupation numbers approaches linearity.\cite{phys_rev_let_2008_piecewise_linearity_finite_infinite,jcp_2015_piecewise_linearity_solid_state_finite}  Fig.~\ref{MESIE_frac} shows that this effect becomes already apparent for 
relatively short polyacetylene oligomers with up to 16 double bonds. The inset showing the deviation from linearity for each chain length highlights a substantial reduction in MESIE as the chain length increases.
In contrast to the case of (semi-)local and hybrid functionals with no or only a small fraction of HF exchange, this behavior is not observed when using full HF or hybrid functionals that include a very large amount of HF exchange. As soon as the curvature of the energy becomes concave, the localization error becomes effective (see Fig.~\ref{MESIE_frac}), meaning it is favorable for the added/removed electron to stay localized. In this situation, increasing the system size has no effect on the shape of the energy curve.\cite{phys_rev_let_2008_piecewise_linearity_finite_infinite}

The appearance of piecewise linearity for large finite systems and the associated loss of the physical relevance of the IP tuning scheme are responsible for the lack of size consistency and the related size dependence and consequential
failure of IP tuned functionals to accurately describe the structural, electronic, and optical properties polyacetylene oligomers of increasing length. Our aim is to investigate to which extent these problems are carried forward when using IP tuned functionals as a starting point for $G_0W_0$.
As DSLE tuning is also based on enforcing straight line behavior, it can be expected that it would face similar difficulties. A strong indication for this is given by the 
size dependence of the fraction of exact exchange found using DSLE tuning. 
In contrast to the two tuning methods enforcing straight line behavior, $G_0W_0$ shows no size dependence. How this affects its ability to accurately describe the evolution of IP with increasing polyacetylene chain length
remains to be seen. 

\section{Computational Details}
Ground state geometry optimizations of polyacetylene oligomers with N=4 to N=22 (see Fig.~\ref{polyacetylene}) were performed in Gaussian09\cite{g09} using a 6-311G basis set, a (75,302) grid, and PBEh* (45.73\% exact exchange), a functional that was specifically parameterized to give excellent results for the BLA of polyacetylene and its evolution with oligomer length.\cite{jcp_2012_wPBE_PBEh_BLA_MESIE} Butadiene (N=2) was excluded from the calculations, as it has been shown to posses very different properties from the other oligomers due to its shortness.\cite{phys_rev_let_1996_BLA_scaling_bandgap}
The resulting geometries all show $C_{2h}$ symmetry. 
The subsequent calculations were carried out using these geometries with no further structure optimization to ensure accurate BLA.
The total energy curves for fractional electron numbers used to show the appearance of piecewise linearity were carried out in PSI4\cite{psi4} using PBE and HF with a cc-pVTZ basis set.
All tuning was done by adapting $\alpha$ in PBEh, considering two decimal places. Details on all of the tuning procedures can be found in the background section.
IP, $G_0W_0$ and DSLE tunings, as well as $G_0W_0$ corrections and most $\Delta$SCF calculations were carried out using FHI-AIMS\cite{fhi_aims}, with a tier 4 basis set for all calculations including $G_0W_0$ and a 
tier 2 basis set for all others. The $G_0W_0$ integration was done following the default two pole model, a comparison with the converged Pade model for the shortest chainlengths can be found in the supporting information. 
For the cation calculations needed for $\Delta$SCF energies for functionals including a large amount of exact exchange ($G_0W_0$ tuned and HF), an increasing amount of spin contamination with chain length was found using unrestricted Hartree-Fock
(UHF), leading to complete failure to converge for longer chains. In these cases we employed restricted open-shell Hartree-Fock (ROHF) calculations in Gaussian for both the neutral and the cation systems for all chain lengths, using a 6-311G(d) basis set.\\
Basis-set extrapolated CCSD(T) is generally considered to be a reliable reference for vertical IPs.\cite{jctc_2016_accurate_IP_I_ccsdt_reference,GW100,GW_acenes}
Due to the very high computational
cost and memory requirements, however, CCSD(T) IPs could only be calculated for Butadiene (N=2) and Octatetraene (N=4). An alternative method called domain based local pair-natural orbital coupled-cluster (DLPNO-CCSD(T) has recently been developed and improved to obtain linear scaling with system size,
while reproducing the CCSD(T) total energy with very high accuracy.\cite{dlpno1,dlpno2,dlpno3} For the case of the polyacetylene oligomers studied here, we found the deviation of the DLPNO-CCSD(T) IPs from the full, basis-set extrapolated CCSD(T) IPs to be 0.02\% for N=2 and 0.05\% for N=4. Based on this excellent agreement, we have therefore chosen to use DLPNO-CCSD(T) IPs as reference data for all chain lengths up to N=22. 
All CCSD(T) and DLPNO-CCSD(T) calculations were performed using ORCA.\cite{orca} An unrestricted reference wavefunction was used for both the neutral and cationic system to obtain $\Delta$SCF IPs. All calculations were performed within
the resolution of identity combined with a chain of spheres exchange (RIJCOSX)\cite{rijcosx} approximation, using ORCAs automatic basis set limit extrapolation. For CCSD(T) N=2 and DLPNO-CCSD(T) N=2-16 the extrapolation was based on cc-pVTZ/QZ\cite{cc-PVTZ/C}, for CCSD(T) N=4 and DLPNO-CCSD(T) N=22 the extrapolation
was based on cc-pVDZ/TZ due to memory constraints. To refer to this (minor) difference in how the numbers were obtained, the DLPNO-CCSD(T) data point for N=22 has been marked with a star in all figures. The DLPNO-CCSD(T) IPs for all chainlengths can be found in the supporting information. 

\section{Results}
\begin{figure}[tb]
\includegraphics[width=1\linewidth]{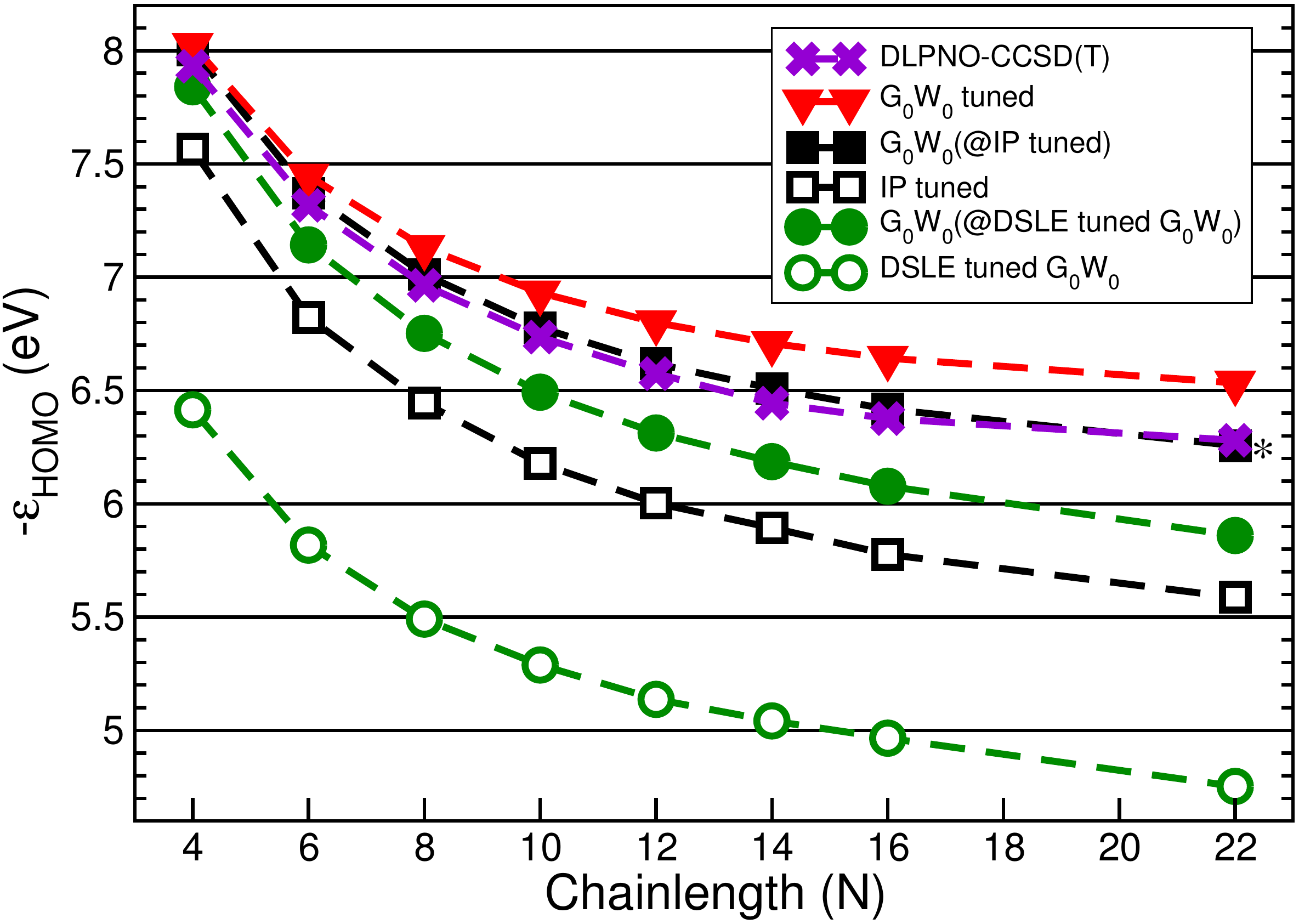}
\caption{Negative HOMO eigenvalues for IP-tuned and $G_0W_0$ tuned PBEh before and after a $G_0W_0$ correction, as well as eigenvalues for DSLE tuned $G_0W_0$ and its underlaying DFT starting point for polyacetylene oligomers of increasing chain length.
DLPNO-CCSD(T) IPs are given as a reference.\label{ip}}
\end{figure}

A summary of the (negative) DFT HOMO eigenvalues and $G_0W_0$ HOMO quasiparticle energies found using each of the tuning methods is provided in Fig.~\ref{ip}, along with the reference data obtained from DLPNO-CCSD(T). For the $G_0W_0$ tuned functional only one set of results is given as, by construction, the DFT and $G_0W_0$ eigenvalues are equal. For short oligomers, all $G_0W_0$ calculated
IPs as well as the HOMO eigenvalue from the IP-tuned functional show good agreement with the DLPNO-CCSD(T) reference. Only the HOMO eigenvalue obtained from the DFT calculation underlaying the DSLE tuning shows poor agreement with the reference data, severely underestimating IPs by more than 1 eV. Given the small fractions of HF exchange included in these functionals (see Fig.~\ref{tuning}), this result is, of course, not surprising. Importantly, the DSLE tuning procedure focuses on reducing the DSLE {\sl after} the $G_0W_0$ correction; the DFT calculation merely provides a starting point and, therefore, is not meant to yield accurate HOMO eigenvalues.

At medium and longer chain lengths, a clear difference between the performance of the different tuning procedures arises. The $G_0W_0$ tuned IPs fall off too slowly, resulting in overestimated IPs with an error increasing with chain length. While the DSLE tuned $G_0W_0$ results show good agreement with the reference for very short chains, the IPs fall off too fast and an increasing underestimation of the IP with increasing chain length can be observed.
IP tuning by itself underestimates the IPs, even for the shorter chain lengths, and slopes off too fast, resulting in an increasing deviation from the reference data for longer chains. This result is a direct consequence of the size dependence of the IP-tuning procedure and resembles the results previously obtained for the BLA\cite{jcp_2012_wPBE_PBEh_BLA_MESIE} and optical band gaps\cite{phys_rev_let_2016_tuning_lrc_polyene_baer_paper} of polyacetylene. 
This underlines once again that, while IP-tuning has been demonstrated to yield excellent results for the IPs and EAs of small to medium-sized organic $\pi$-conjugated molecules, great care has to be taken when applying it to large, conjugated molecules, in particular in the context of the oligomer approach. Surprisingly, however, the IPs obtained from applying a $G_0W_0$ correction to the IP-tuned eigenvalues reproduce the reference data extremely accurately, showing both good agreement for absolute values as well as the slope with increasing chain length.
\begin{figure}[h!]
\includegraphics[width=1\linewidth]{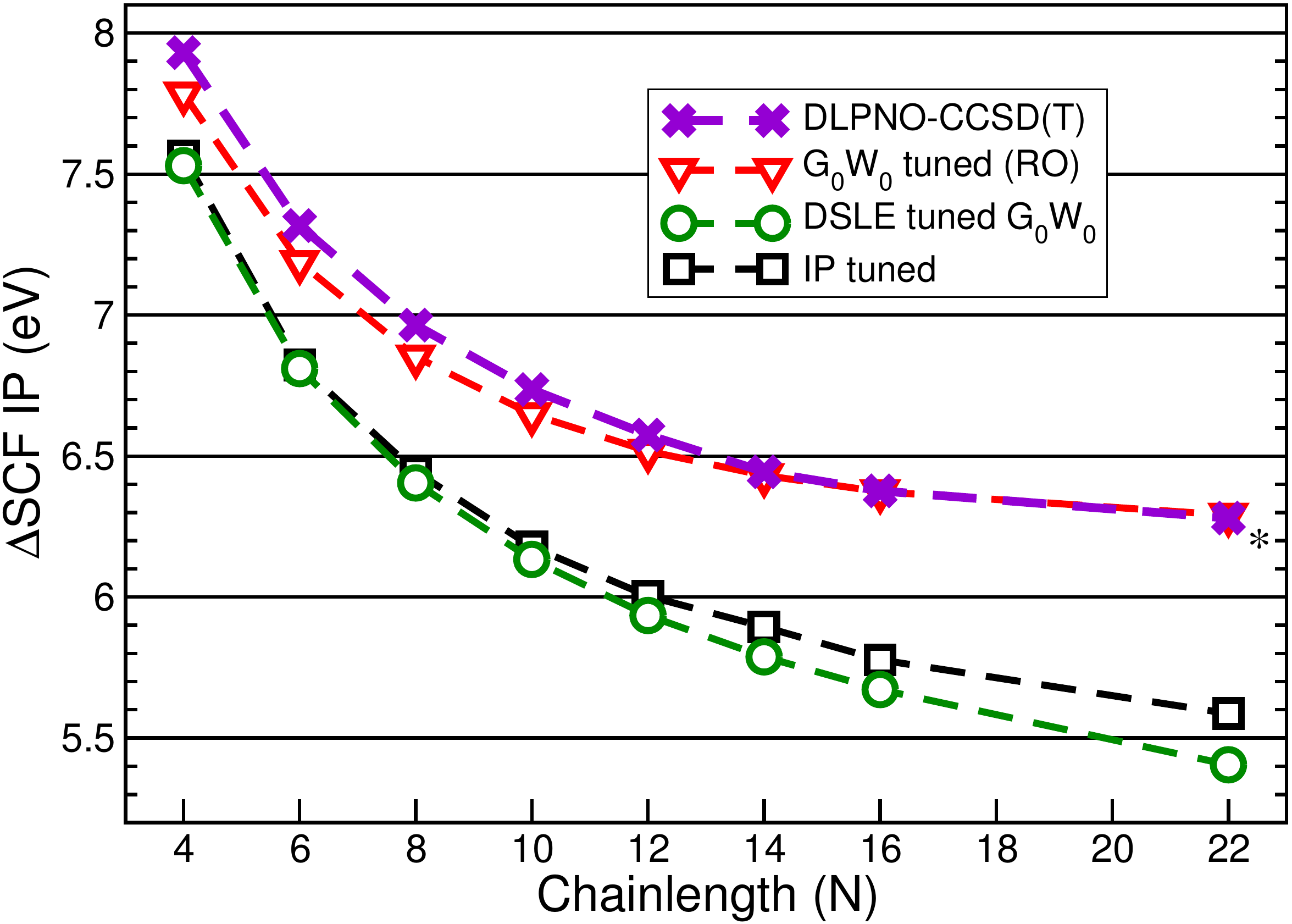}
\caption{$\Delta$SCF IPs for IP-tuned, $G_0W_0$ tuned and DSLE tuned PBEh for polyacetylene oligomers of increasing chainlength. DLPNO-CCSD(T) IPs are given as a reference.\label{delta_scf}}
\end{figure}
\hfill \\
IPs can also be calculated as the difference in total energy between the neutral and cationic system, known as the $\Delta$SCF approach. 
While this should be equal to the HOMO eigenvalue for the exact functional according to Eq.~(\ref{eq:ip_theorem}), there can be a significant difference between the HOMO eigenvalue and the $\Delta$SCF IP for approximate functionals. For the infinite chain limit, however, the HOMO eigenvalue equals its corresponding $\Delta$SCF IP for all functionals that show a delocalization error, as discussed above. Therefore, it is interesting for us to also study the evolution of the $\Delta$SCF IP with increasing chain length. These results are provided in Fig.~\ref{delta_scf}.\cite{note2}  

IP-tuned and DSLE tuned functionals lead to very similar $\Delta$SCF IPs, both underestimating the reference data and falling off too rapidly with increasing chain length. This error can be traced back to their size dependence: The strong decrease in exact exchange with increasing chain length leads to increasingly inaccurate $\Delta$SCF IPs. Since the the IP-tuning procedure forces the HOMO eigenvalue to equal the $\Delta$SCF IPs, this result further illustrates why the HOMO IP obtained from IP-tuned functionals becomes more and more inaccurate for $\pi$-conjugated molecular chains of increasing length.
In contrast to IP- and DSLE tuning, $G_0W_0$ tuning results in a very good agreement with the reference, with only minor deviations from both the absolute IP values and their slope. We attribute this to the approximate size-consistency of the $G_0W_0$ tuning procedure, resulting in an almost constant amount of exact exchange and, thus, allowing the $\Delta$SCF IPs to evolve correctly with increasing chain length. 

\begin{figure*}[tbh]
\includegraphics[width=0.9\linewidth]{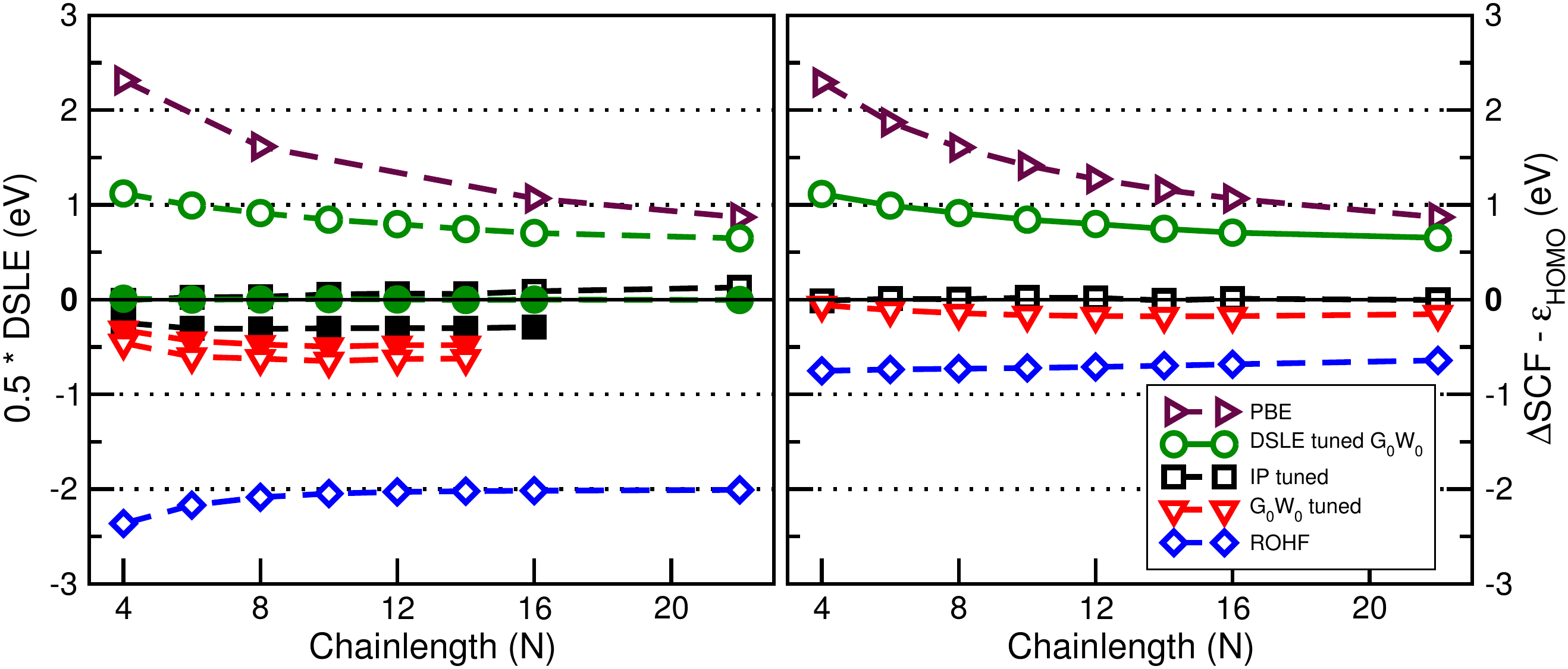}
\caption{0.5 * DSLE (left) and difference between $\Delta$SCF and eigenvalue IPs (right) for  IP-tuned, $G_0W_0$ tuned and DSLE tuned PBEh, as well as PBE and HF for polyacetylene oligomers of increasing chain length. The unfilled
symbols represent the data for the underlaying DFT calculations, while the filled symbols show the data after a $G_0W_0$ correction.\label{dsle}}
\end{figure*}
\section{Further Discussion}
To interpret the different tuning methods' behavior with increasing chain length and to highlight their respective strengths and shortcomings, it is helpful to study both their DSLE and the difference between the $\Delta$SCF IP
and the negative HOMO eigenvalues, as shown in Fig.~\ref{dsle} on the left and right side, respectively. These two values represent the errors minimized by DSLE and IP tuning, respectively. 
A clearer comparison between the two quite similar errors can be drawn following the approach by Bruneval\cite{PhysRevLett.103.176403} and plotting half of the DSLE as described by equation \ref{eq:dsle_general}.
This follows from the fact that the
energy curve for fractional particle numbers can be expanded as a second order polynomial and hence the error can be expressed as the mean value of HOMO and LUMO eigenvalues. 

PBE and HF have been included to provide the limiting cases
of 0\% HF and 100\% HF, as all our tuned functionals are a combination of these two. PBE is a semi-local functional, with a strong delocalization error, represented by a positive DSLE and a $\Delta$SCF IP significantly larger than
the negative HOMO eigenvalue. Both errors, however, show a strong decrease with chain length. 
This behavior of PBE can be  rationalized by the appearance of piecewise linearity with increasing system size demonstrated in Fig.~\ref{MESIE_frac}.
HF on the other hand shows a large negative DSLE and $\Delta$SCF IP smaller than the negative HOMO eigenvalue, with almost no change in error with increasing chain length. This shows the localization
error, which prevents the appearance of piecewise linearity for increasing system size, hence leading to a consistent performance of the functional for all oligomer lengths. 
The difference between the 0.5 * DSLE and the $(\Delta$SCF - $\epsilon_{\mathrm{HOMO}})$ seen in figure \ref{dsle} can be explained by the behavior of the cation LUMO in the ROHF approach, as can be seen in the supporting information. 

The largest fraction of exact exchange is found for the $G_0W_0$ tuned functional, where $\alpha\approx 0.81$. Fig.~\ref{dsle} shows a similiar behavior of the $G_0W_0$ tuned and HF methods, with both approaches displaying a 
negative DSLE, a $\Delta$SCF IP smaller than the negative HOMO eigenvalue and almost constant errors with chain length. It follows that $G_0W_0$ tuned functionals exhibit a localization error.
This is also reflected in the density difference of the neutral and cationic systems obtained from the $G_0W_0$ tuned functional when compared to the DLPNO-CCSD(T) result (see supporting information). The tendency of $G_0W_0$ to exhibit a localization error for DFT starting points with a large amount of exact exchange has been previously discussed in literature.\cite{phys_rev_b_2016_dsle_tuning}

The small discrepancies found for the DSLE of the $G_0W_0$ tuned functional before and after the $G_0W_0$ correction can be traced back to a change in the cation LUMO, as only the HOMO is tuned to give identical DFT and $G_0W_0$ HOMO energies. A figure showing how the $G_0W_0$ correction affects the 5 highest occupied and 5 lowest unoccupied orbitals of the neutral system can be found in the supporting information. 

\begin{figure}[tbh]
\includegraphics[width=1\linewidth]{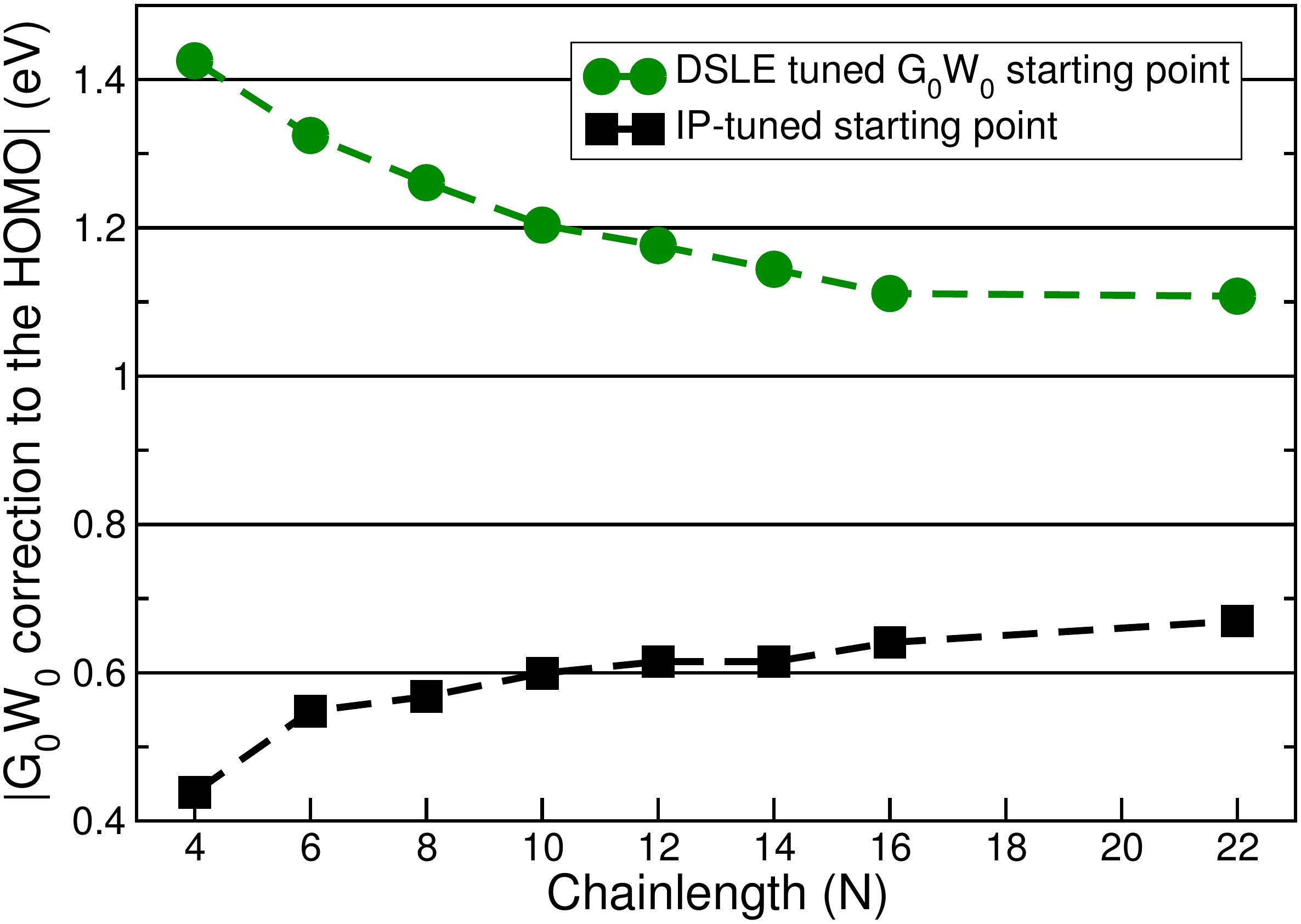}
\caption{The $G_0W_0$ correction to the HOMO eigenvalue for IP-tuned and DSLE tuned PBEh with increasing chainlength of polyacetylene.\label{gw_corr}}
\end{figure}

DSLE tuning leads to the smallest fractions of exact exchange (0.2 $<\alpha<$ 0.4), and the corresponding DFT starting point shows a behavior similar to PBE. Again, a strong delocalization error is apparent through a positive DSLE and a $\Delta$SCF IP much 
larger than the negative HOMO eigenvalue. Again, this delocalization error can be directly observed in the density difference between the neutral and cationic systems (see supporting information).  Even though a small amount of exact exchange is present, the appearance of piecewise linearity for long chains can be observed, the errors seemingly decreasing with chain length. By construction, the tuning 
minimizes the DSLE after the $G_0W_0$ correction, which is clearly seen in Fig.~\ref{dsle} (left). As the DFT starting point appears to have decreasing DSLE with increasing chain length, however, the correction needed from $G_0W_0$ 
to enforce linearity decreases (see Fig.~\ref{gw_corr}). 

For IP tuned functionals linearity is already enforced approximately in the DFT starting point, and the $\Delta$SCF and eigenvalue IPs are equal by construction. The DSLE is almost zero, showing a slight deviation with increasing
chain length caused by the behavior of the cation LUMO (see supporting information for details). Applying the $G_0W_0$ correction on top of IP tuned functionals leads to a slightly negative DSLE indicating a localization error. 
However, the $G_0W_0$ correction increases with chain length,
shown in Fig.~\ref{gw_corr}, thus working against the error in the DFT starting point introduced by the appearance of piecewise linearity.

\section{Conclusions}
In summary, we conclude that the ability of $G_0W_0$ calculations to accurately describe the evolution of the HOMO IP with chain length for $\pi$-conjugated molecular chains such as polyacetylene relies on a number of factors:

Firstly, the amount of exact exchange present in a hybrid functional determines the slope of the IP with increasing chain length, with low fractions of exact exchange causing the IPs to fall off too rapidly and high
fractions of exact exchange causing the IPs to not fall off fast enough. This corresponds to similar trends observable for other properties such as BLA and excitation energies.\cite{jcp_2011_range_sep_conj,jctc_2016_BLA_Ex_thermal} Eliminating the localization/delocalization 
error in either the DFT or the $G_0W_0$ step (as resulting from IP and DSLE tuning, respectively) does not guarantee accurate IPs or a correct evolution with chain length.

Secondly, $G_0W_0$ tuning leads to a slight localization error, observable in the eigenvalue IPs. In contrast to the other methods it is, however, not strongly size dependent and hence leads to $\Delta$SCF IPs in good agreement 
with the reference data.

Finally, a $G_0W_0$ correction improves the DFT HOMO IPs for all tuning methods. The best agreement with the CCSD(T) reference data, however, is found when using an IP tuned starting point, yielding highly accurate IPs for all oligomer sizes. Given the strong size dependence of the IP-tuned starting point and, related to that, the decreasing accuracy of the IP-tuned HOMO eigenvalues for extended $\pi$-conjugated systems, this result could not be expected. However, it seems that the
$G_0W_0$ correction can counteract the error introduced by the size dependence, at least for the example of polyacetylene studied in this work. This is likely to be caused by a favourable cancelation of errors between the IP-tuned starting point and the $GW$ approximation. 
We conclude that, although there are known shortcomings of the IP-tuning approach for large $\pi$-conjugated molecules due to its size dependence, this apparently does not diminish the suitability of such functionals as a good starting point for $G_0W_0$ calculations on large, $\pi$-conjugated molecules.

\begin{suppinfo}
The supporting information includes the IPs obtained from the DLPNO-CCSD(T) calculations, as well as additional data showing the difference in DSLE obtained from fractional occupation curves and HOMO/LUMO eigenvalues, and data comparing the $G_0W_0$ eigenvalues of the HOMO and LUMO
for N=4 and 6 found following the two pole model for $GW$ integration compared to the converged Pade results.  
Also included are figures showing
the $G_0W_0$ correction to the five highest occupied as well as the five lowest unoccupied molecular orbitals following the 
$G_0W_0$ tuning procedure, and the straight
line error of the cation LUMO and neutral system HOMO for ROHF and after IP tuning. Finally, density plots for the difference between the neutral and the cationic system as well as for the neutral HOMO are included for all tuning 
methods and the DLPNO-CCSD(T) reference. 
\end{suppinfo}

\begin{acknowledgement}
The authors thank L. Gallandi 
and P. Rinke
for helpful discussions.
\end{acknowledgement}

\bibliography{References_3}

\begin{mcitethebibliography}{89}
\bibitem[not()]{notevxc}
In a standard Kohn-Sham formulation, the exchange-correlation operator
  $\hat{v}_{\mathrm{xc}}$ is a local, multiplicative potential. In a
  generalized Kohn-Sham calculation, e.g., using a hybrid functional, the
  exchange-part of $\hat{v}_{\mathrm{xc}}$ becomes a non-local operator.\relax
\mciteBstWouldAddEndPunctfalse
\mciteSetBstMidEndSepPunct{\mcitedefaultmidpunct}
{}{\mcitedefaultseppunct}\relax
\EndOfBibitem
\bibitem[not()]{note2}
$\Delta$SCF IPs are only available for DFT calculations, as the single point
  $G_0W_0$ correction does not lead to a new total energy.\relax
\mciteBstWouldAddEndPunctfalse
\mciteSetBstMidEndSepPunct{\mcitedefaultmidpunct}
{}{\mcitedefaultseppunct}\relax
\EndOfBibitem
\end{mcitethebibliography}
\end{document}